\documentclass[printer]{aa}

\usepackage{graphicx}
\usepackage{txfonts}
\begin{document} 

\title{The VLT/NaCo large program to probe the occurrence of exoplanets and brown dwarfs at wide orbits\thanks{Based on observations collected at the European Southern Observatory, Chile (ESO Large Program 184.C-0157 and Open Time 089.C-0137A and 090.C-0252A).} }

   \subtitle{III. The frequency of brown dwarfs and giant planets as companions to solar-type stars}

 \titlerunning{The frequency of brown dwarfs and giant planets as companions to solar-type stars}

\author{M. Reggiani \inst{\ref{inst1},\ref{inst2}}
\and M. R. Meyer \inst{\ref{inst1}} 
\and G. Chauvin \inst{\ref{inst3}}
\and A. Vigan \inst{\ref{inst4}}
\and S. P. Quanz \inst{\ref{inst1}} 
\and B. Biller \inst{\ref{inst5}}
\and M. Bonavita \inst{\ref{inst5},\ref{inst6}}
\and S. Desidera\inst{\ref{inst6}}
\and P. Delorme \inst{\ref{inst3}}
\and J. Hagelberg \inst{\ref{inst7},\ref{inst17}}
\and A.-L. Maire \inst{\ref{inst6}}
\and A. Boccaletti \inst{\ref{inst8}}
\and J.-L. Beuzit \inst{\ref{inst3}}
\and E. Buenzli\inst{\ref{inst1}}
\and J. Carson\inst{\ref{inst10}}
\and E. Covino\inst{\ref{inst11}}
\and M. Feldt\inst{\ref{inst9}}
\and J. Girard \inst{\ref{inst12}}
\and R. Gratton \inst{\ref{inst6}}
\and T. Henning\inst{\ref{inst9}}
\and M. Kasper \inst{\ref{inst13}}
\and A.-M. Lagrange \inst{\ref{inst3}}
\and D. Mesa\inst{\ref{inst6}}
\and S. Messina \inst{\ref{inst14}}
\and G. Montagnier \inst{\ref{inst4}}
\and C. Mordasini\inst{\ref{inst15}}
\and D. Mouillet \inst{\ref{inst3}}
\and J. E. Schlieder  \inst{\ref{inst16}}
\and D. Segransan  \inst{\ref{inst17}}
\and C. Thalmann \inst{\ref{inst1}} 
\and A. Zurlo \inst{\ref{inst18},\ref{inst19}}
          }

   \institute{Institute for Astronomy (IfA), ETH, Zurich\email{reggiani@phys.ethz.ch}\label{inst1} 
   \and D\'epartement d'Astrophysique, G\'eophysique et Oc\'eanographie, Universit\'e de Li\`ege, 17 All\'ee du Six Ao\^ut, 4000 Li\`ege, Belgium\label{inst2}
   \and UJF-Grenoble1/CNRS-INSU, Institut de Plan\'{e}tologie et d’Astrophysique de Grenoble UMR 5274, 38041 Grenoble, France\label{inst3}
   \and   Aix-Marseille Universit\'{e}, CNRS, LAM (Laboratoire d’Astrophysique de Marseille) UMR 7326, 13388 Marseille, France\label{inst4}
   \and Institute for Astronomy, University of Edinburgh, Blackford Hill, Edinburgh EH9 3HJ, UK\label{inst5}
   \and INAF – Osservatorio Astronomico di Padova, Vicolo dell Osservatorio 5, 35122 Padova, Italy\label{inst6}
   \and Institute for Astronomy, University of Hawaii, 2680 Woodlawn Drive, Honolulu, HI 96822 USA \label{inst7}
   \and LESIA, Observatoire de Paris Meudon, 5 Pl. J. Janssen, 92195 Meudon, France  Italy\label{inst8}
   \and Max-Planck Institute for Astronomy, K\"{o}nigstuhl 17, 69117 Heidelberg, Germany \label{inst9}
   \and Department of Physics and Astronomy, College of Charleston, Charleston, SC 29424 , USA \label{inst10}
   \and INAF Osservatorio Astronomico di Capodimonte via Moiarello 16, 80131 Napoli, Italy \label{inst11}
   \and European Southern Observatory, Casilla 19001, Santiago 19, Chile \label{inst12}
   \and European Southern Observatory, Karl Schwarzschild St, 2, 85748 Garching, Germany \label{inst13}
   \and  INAF – Catania Astrophysical Observatory, via S. So a 78, 95123 Catania, Italy\label{inst14}
   \and Physikalisches Institut, University of Bern, Sidlerstrasse 5, CH-3012 Bern, Switzerland \label{inst15}
   \and NASA Ames Research Center, Moffett Field, CA, USA \label{inst16}
   \and Geneva Observatory, University of Geneva, Chemin des Mailettes 51, 1290 Versoix, Switzerland \label{inst17}
\and N\'ucleo de Astronom\'ia, Facultad de Ingenier\'ia, Universidad Diego Portales, Av. Ejercito 441, Santiago, Chile \label{inst18}
\and Millennium Nucleus ``Protoplanetary Disk'', Departamento de Astronom\'ia, Universidad de Chile, Casilla 36-D, Santiago, Chile  \label{inst19} }

   \date{---}

 
  \abstract
   {In recent years there have been many attempts to characterize the occurrence and distribution of stellar, brown dwarf (BD) and planetary-mass companions to solar-type stars, with the aim of constraining formation mechanisms. 
   From radial velocity observations a dearth of companions with masses between 10-40 $M_{Jupiter}$ has been noticed at close separations, suggesting the possibility of a distinct formation mechanism for objects above and below this range.} 
   {We present a model for the substellar companion mass function (CMF). It consists of the superposition of the planet and BD companion mass distributions, assuming that we can extrapolate the radial velocity measured companion mass function for planets to larger separations and the stellar companion mass-ratio distribution over all separations into the BD mass regime. By using both the results of the VLT/NaCo large program (NaCo-LP, P.I. J. L. Beuzit) and the complementary archive datasets that probe the occurrence of planets and BDs on wide orbits around solar-type stars, we place some constraints on the planet and BD distributions.}
   {We developed a Monte Carlo simulation tool to predict the outcome of a given survey, depending on the shape of the orbital parameter distributions (mass, semi-major axis, eccentricity and inclination). Comparing the predictions with the results of the observations, we calculate how likely different models are and which can be ruled out.}
   {Current observations are consistent with the proposed model for the CMF, as long as a sufficiently small  outer truncation radius ($\lesssim$100 AU) is introduced for the planet separation distribution. Some regions of parameter space can be excluded by the observations.}
   {We conclude that the results of the direct imaging surveys searching for substellar companions around Sun-like stars are consistent with a combined substellar mass spectrum of planets and BDs. This mass distribution has a minimum between 10 and 50 $M_{Jupiter}$, in agreement with radial velocity measurements. In this picture the dearth of objects in this mass range would naturally arise from the shape of the mass distribution, without the introduction of any distinct formation mechanism for BDs. Such a model for the CMF allows to determine what is the probability for a substellar companion as a function of mass to have formed in a disk or from protostellar core fragmentation, as such mechanisms overlap in this mass range.}


   \maketitle
%
\section{Introduction}\label{sect:3.1}
Binary systems have been observed and characterized for almost 100 years \citep[see e.g. ][]{Aitken1935}.
Since the seminal works by \cite{Duquennoy1991} and \cite{Fischer1992}, the properties of stellar binaries have also been widely studied. One of the most interesting parameters of a binary system is the mass-ratio $q=M_{2}/M_{1}$, defined as the ratio of the secondary ($M_{2}$) over the primary mass ($M_{1}$). The distribution of $q$ values for a sample of binaries is the companion mass-ratio distribution (CMRD). Several surveys in the past decades focused on the detection of stellar binaries with the purpose of characterizing the occurrence of companions and their mass distribution both in the field \citep[e.g.][]{Raghavan2010, Janson2012} and in star forming regions \citep[e.g.][]{Patience2002}. \cite{Reggiani2013}, as an update of \cite{Reggiani2011}, have shown that in the field the CMRD is consistent with being universal, independent of primary mass and separation in the range covered by the observations, and can be fit by a single power-law slope $dN/dq \propto q^{\beta}$, with $\beta=0.25\pm0.29$. In addition, N-body simulations suggest that the CMRD is only modestly affected by dynamics, even in dense clusters, as opposed to the semi-major axis (SMA) distribution \citep{Parker2013}. The CMRD therefore seems to be a good diagnostic for different star formation mechanisms.

While the distributions of masses and orbital parameters are quite well established for stellar companions, the shape of the companion mass function (CMF) and of the SMA distribution in the substellar and planetary mass regime is still poorly understood.  The over 500 extrasolar planets discovered with the radial velocity (RV) method allowed astronomers to fit power-law slopes to the mass and semi-major axis distributions for planets with masses between 0.3-10 $M_{Jupiter}$ and within $\sim$3 AU \citep[see e.g.][]{Fischer2005,Cumming2008}. However, nothing guarantees that the same behavior holds at larger separations or higher masses. Direct imaging is a detection method that allows us to characterize large-separation exoplanets. In the past decade many direct imaging surveys \citep[e.g.][]{Lowrance2005,Biller2007,Lafreniere2007,Chauvin2010,Heinze2010,Vigan2012,Biller2013,Rameau2013,Wahhaj2013,Chauvin2015,Brandt2014a,Bowler2015} have been carried out to evaluate the occurrence of giant planets in wide orbits (10-500 AU). Besides a few planets \citep[e.g.][]{Kuzuhara2013,Rameau2013b} and brown dwarfs (BDs) \citep[e.g.][]{Chauvin2005,Biller2010,Mugrauer2010,Wahhaj2011,Carson2013} detected, many surveys found no planetary companions. The null results cannot be used to fit power-laws to the planetary distributions beyond 10 AU. However they suggest a truncation of the planet semi-major axis distribution at a few tens of AU in order to reproduce the RV statistics below 3 AU (assuming the same power-law slope holds). 

Regarding the substellar mass range, BDs were originally proposed as a separate class of objects, with intermediate masses between stars and planets. High contrast observations have unambiguously revealed the presence of substellar objects, as companions to nearby stars, as well as BD-BD systems \citep[e.g.][]{Chauvin2005,Dupuy2011}. Some were also found in isolation in the field \citep[e.g.][]{Burgasser2003}. Their existence indicates that the formation mechanisms proposed to form stars (turbulent fragmentation, collapse and fragmentation, disk fragmentation) can actually form objects down to a few Jupiter masses.
Star and planet formation mechanisms therefore overlap in the planetary-mass regime. Companions in this mass range could be the lower mass tail of the stellar CMRD as well as the higher mass end of the planet CMF, but they have usually been excluded by the statistical analyses of direct imaging surveys.  Different features in the frequency of BD companions and planets as a function of mass and semi-major axis could help distinguish different formation mechanisms \citep[see also][]{Brandt2014b}. Although spectroscopy may help in investigating whether a given detected object formed by one process or another, large surveys could contribute in assessing whether a stellar-like or planet-like formation process dominates the CMF in a given companion mass range.

As a project within the VLT/NaCo large program \citep[hearafter NaCo-LP, P.I. J. Beuzit,][]{Desidera2015,Chauvin2015}, the work presented in this paper represents the first attempt to place some constraints on the full CMF in the substellar regime. It takes into account both BDs and planets and makes use of the results of most of the direct imaging surveys of solar-type primaries currently available.
In Sections~\ref{sect:3.2} and~\ref{sect:3.3} we present our model for the mass distribution of substellar companions and the Monte Carlo simulation tool that we developed to test it.
Then, we describe the datasets that we adopted for the analysis (Section~\ref{sect:3.4}) and the results of the Monte Carlo simulations based on these data (Section~\ref{sect:3.5}). 
Finally, we discuss the results in Section~\ref{sect:3.6} and summarize the conclusions in Section~\ref{sect:3.7}.


\begin{table}[!h]
\caption{\label{table:3.1} Planet and BD distributions: model values.}
\centering
\begin{tabular}{cc}
\hline\hline
Parameter & Values\\
\hline
    $\alpha_{BD}$ & 1.25\tablefootmark{a} \\
    $\alpha_{pl}$ & -0.31\tablefootmark{b} \\    
    $a_{c_{BD}}$ & 50 AU\tablefootmark{c}\\ 
    $\sigma_{BD}$ & 1.68\tablefootmark{c} \\ 
    $\beta_{pl}$ &  0.39\tablefootmark{b}\\ 
    $f_{BD}$ &  0.61  for  [q=0.08-1] \& all separations\tablefootmark{c} \\ 
    $f_{pl}$ &  0.0329 for [1-13 $M_{J}$] \&  [0.3-2.5 AU]\tablefootmark{d}\\
        m$_{min_{BD}}$ & 5 M$_{Jupiter}$  \\
        m$_{max_{BD}}$ & 80 M$_{Jupiter}$  \\
        a$_{min_{BD}}$ & 0.1 AU  \\
        a$_{max_{BD}}$ & 10000 AU  \\        
        m$_{min_{pl}}$ & --- \\
        m$_{max_{pl}}$ & 0.1 x M$_{star}$  \\
        a$_{min_{pl}}$ & ---  \\
        a$_{max_{pl}}$ & r$_{cutoff}$\tablefootmark{e} \\
\hline
\end{tabular}
\tablefoot{ 
\tablefoottext{a}{\cite{Reggiani2013}, measured for stellar companions but extrapolated to the BD regime.}
\tablefoottext{b}{\cite{Heinze2010}, based on \cite{Cumming2008} and extrapolated to larger separations.}
\tablefoottext{c}{\cite{Raghavan2010}, measured for stellar companions but extrapolated to the BD regime.}
\tablefoottext{d}{\cite{Heinze2010}, based on \cite{Fischer2005}.}
\tablefoottext{e}{It is a free parameter in our analysis.}
}
\end{table}

\section{Model for the substellar CMF}\label{sect:3.2}
Our model for the CMF is based on the hypothesis that both BDs and planets contribute to the substellar mass distribution. 
In this context, we consider as ``BDs'' all objects that, having formed through a ``stellar-like mechanism'', constitute the lower mass tail of the CMRD, assuming that it can be extrapolated into the BD regime as suggested by \cite{Metchev2009}. We define as ``planets'', instead, all companions that formed in ``standard planet-formation scenarios'' and gave rise to the RV measured CMF but extrapolated to larger separations. There are of course some caveats to these assumptions. For instance, we do not take into account how gravitational instability and core accretion differ in the distributions of orbital parameters of planets. The fact that there are multiple binary formation mechanisms and that both planets and BDs migrate makes everything more complicated. In this paper we  propose a simple, but new, model and test whether it works.  

Instead of trying to fit the substellar CMF with a single power law distribution \citep[see e.g.,][]{Brandt2014b}, we assume the overall frequency of substellar companions to be the sum of two contributions:
\begin{eqnarray}
d^2N_{BD} = C_{BD} \, q^{\alpha_{BD}} \, d\log q \, e^{-\frac{(\log a-\log a_{c_{BD}})^2}{2\,\sigma_{BD}^2}} \, d\log a, \label{eq:1}\\ 
d^2N_{pl}= C_{pl} \, m^{\alpha_{pl}} \, d\log m \, a^{\beta_{pl} } \,d \log a \label{eq:2}
\end{eqnarray}
for BDs and planets, indicated with the ``BD '' and ``pl'' subscripts, respectively. $q$ is the mass ratio, as defined in Section \ref{sect:3.1}. $\alpha_{BD,pl}$ and $\beta_{pl}$ are the exponents of the power-law mass and semi-major axis distributions \citep[c.f.][]{Cumming2008}, whereas $a_{c_{BD}}$ and $\sigma_{BD}$ are the mean and the standard deviation of the BD separation distribution, assuming it is log-normally distributed as for solar type primaries \citep[see e.g.][]{Raghavan2010}.    
$C_{BD}$ and $C_{pl}$ are normalization constants that can be obtained from measurements of the BD and planet frequency $f_{BD,pl}$ over a range of masses and semi-major axes.

In this paper we assume values for the parameters that we think are appropriate for solar-type primaries, typical of the dataset that we use to test this model. 
Our choices are presented in Table~\ref{table:3.1}. Some of these values are measurement based, and some are extrapolations (e.g. the CMRD in the BD regime). Notes in Table~\ref{table:3.1} explain the extrapolations that have been made. The brown dwarf distribution is normalized to the companion frequency, $f_{BD}$, measured by \cite{Raghavan2010} for stellar companions. If extrapolated to the BD mass regime ($q=0.012-0.072$), it gives $f_{BD}=1^{+1}_{-0.6}$\% between 28-1590 AU, which is in agreement with the brown dwarf companion fraction presented by \cite{Metchev2009}, $f_{BD}=3.2^{+3.1}_{-2.7}$\%, over the same mass and separation ranges.\\
\indent Moreover, several physical mechanisms or observational constraints place upper and lower limits to the maximum and minimum mass (m$_{max_{BD,pl}}$, m$_{min_{BD,pl}}$) and separation (a$_{max_{BD,pl}}$, a$_{min_{BD,pl}}$) for these distributions.
For our analysis we assume:
\begin{itemize}
\item the minimum (m$_{min_{BD}}$)  and maximum  (m$_{max_{BD}}$) masses for the BD mass distribution are given by the opacity limit for fragmentation \citep{Low1976} and the hydrogen burning limit  \citep[e.g.][]{Burrows2001},
\item the minimum (a$_{min_{BD}}$) and maximum (a$_{min_{BD}}$) semi-major axes for the BD distribution can be set in order to exclude less than 1\% of companions from the measured stellar separation distribution \citep[e.g.][]{Raghavan2010},
\item the maximum mass (m$_{max_{pl}}$) for the planet mass distribution is constrained by the disk mass (e.g. $\leq$10\% of the mass of the star),
\item the maximum separation (a$_{max_{pl}}$) for the planet separation distribution is possibly set by an outer truncation radius r$_{cutoff}$ (as suggested by previous direct imaging surveys), as long as planet-planet scattering does not affect significantly the distribution. Such a truncation radius is treated as a free parameter in our analysis.\\
\end{itemize} 

\begin{figure}
   \centering
   \resizebox{\hsize}{!}{\includegraphics{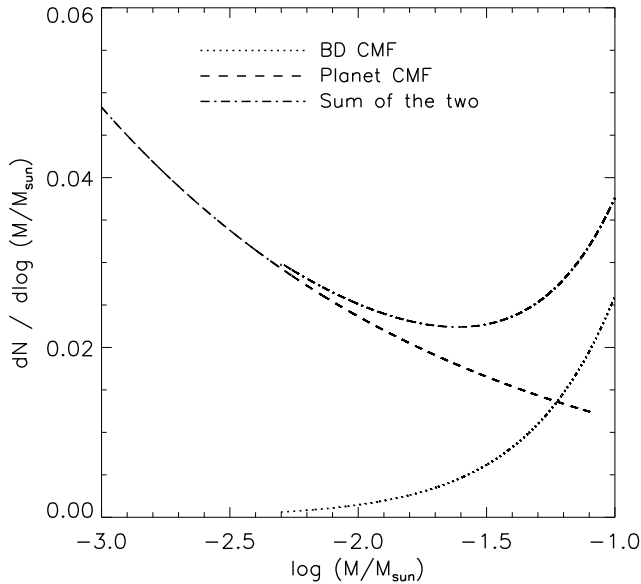}}
      \caption[Combined substellar CMF]{Combined substellar CMF. The discontinuity at $\log M/M_{\odot}$=-2.3 is due to the opacity limit for fragmentation that set a minimum mass for objects formed through core fragmentation. The minimum mass is assumed to be 5 $M_{Jupiter}$ in our model. }
         \label{figure:3.1}
   \end{figure}

The combined CMF in the substellar regime for this choice of parameters and assuming a 1 $M_{\odot}$ star is shown in Figure~\ref{figure:3.1}.
We have assumed that the opacity limit for fragmentation sets a lower limit to the BD mass distribution, creating a discontinuity in the CMF. The opacity limit for fragmentation occurs when the gravitational potential energy that is released during the collapse of a molecular cloud core exceeds the energy that can be radiated away \citep{Low1976}. Theories of core fragmentation predict that fragmentation cannot take place at densities higher than 10$^{-13}$ g cm$^{-3}$, corresponding to a minimum mass of a few Jupiter masses \citep[e.g. 1-10 $M_{Jupiter}$,][]{Low1976,Silk1977,Boss2001}. A challenging observational goal would be to test this aspect of our model. In this work we assumed a minimum mass for BDs of 5 $M_{Jupiter}$.  

We note that the analysis presented in this paper does not take into account the uncertainties on the model parameters described in Table~\ref{table:3.1}, as this would be computationally too expensive. This aspect can certainly be improved in the future. Furthermore better constraints on the BD and planet orbital parameter distributions from observations, as well as predictions from core accretion or gravitational instability theories will allow us to further test the model.

Previous studies have shown a dependence of the planet properties (e.g. frequency) on stellar properties \citep[e.g. mass and metallicity, ][]{Johnson2010}, however the analysis of the influence of the star properties on this model goes beyond the scope of the paper as our sample is restricted to a narrow range of primary masses. Future observations will allow us to study the dependence of the combined substellar CMF as a function of stellar mass.

\section{Methodology}\label{sect:3.3}

In this section we present the methodology that we developed to test the model for the substellar CMF presented in Section~\ref{sect:3.2}. 
We created a Monte Carlo simulation tool with the aim of predicting the frequency of planets and BDs around a given list of targets and the probability of detection in a given survey, assuming the combined CMF.
The probability of existence of substellar companions around a star depends on their mass and semi-major axis distributions according to our model. The probability of detecting them instead depends on the parameters of the companions (mass, separation, eccentricity, inclination), on the properties of the star (age, distance, mass), and on the sensitivity (and contrast limits) of the instrument used.

The first step in our methodology is therefore to calculate the expected number of existing companions, both BDs and planets, per star.
This quantity is generally expressed as the expectation value of the number of companions per star in a given range of masses and separations:
\begin{eqnarray}
P_{BD}= C_{BD} \, \int_{q_{min_{BD}}}^{q_{max_{BD}}}q^{\alpha_{BD}} \, d\log q \, \int_{a_{min_{BD}}}^{a_{max_{BD}}} \, e^{-\frac{(\log a-a_{c_{BD)}})^2}{2 \sigma_{BD}^2}} \, d\log a,\label{eq:3}\\
P_{pl}=C_{pl} \, \int_{m_{min_{pl}}}^{m_{max_{pl}}}m^{\alpha_{pl}} \, d\log m \, \int_{a_{min_{pl}}}^{a_{max_{pl}}} \, a^{\beta_{pl}} \, d\log a, \label{eq:4}
\end{eqnarray}

\noindent where the integral limits indicate the mass/mass ratios and semi-major axis ranges of interest for BDs (``BD'' subscript) and planets (``pl'' subscript). q$_{min_{BD}}$ and q$_{max_{BD}}$ correspond to the mass ratios for m$_{min_{BD}}$ and m$_{max_{BD}}$, respectively. The other parameters are described in Section~\ref{sect:3.2}.

Once we have calculated the average number of planets and BDs per star in the mass and semi-major axis ranges that we are interested in, we run N simulations of the survey, in which every target star is randomly assigned a number of planets and a number of BDs, based on Poisson statistics with mean $P_{pl}$ and $P_{BD}$ for planets and BDs, respectively. 
In case the Poisson probability returns a number smaller than 0.5, the number of planets or BDs for the given star is zero. However, if the star turns out to have one or more planets (or BDs), the mass and semi-major axis for each one of them is randomly selected from the input planet (or BD) distributions (see Section~\ref{sect:3.2}). 
The eccentricity is randomly selected from the \cite{Juric2008} distribution, of the form $P(\epsilon)= \epsilon * e^{[-\epsilon^2/2\sigma^2]} $ with $\sigma =0.3$, and an inclination is randomly assigned.
The time spent on the orbit, as a function of orbital location, is explicitly taken into account as a function of orbital parameters when converting the semi-major axis into a projected separation in AU.\\
Once every simulated planet or BD is assigned all its orbital properties, we compare its mass and separation with the sensitivity limits of the survey for a specific target to determine which planets or BDs could have been detected. Generally, the sensitivity limits are given for each star as a contrast curve, meaning apparent magnitude (or contrast) as a function of the angular separation. In order to perform a direct comparison of the simulated properties with the contrast curves, we use a family of substellar evolutionary models \citep[e.g. COND models,][]{Baraffe2003} and the information about distance and age of the stars to transform companion masses and projected separations into apparent magnitudes and angular separations. If the combination of magnitude and separation for a given companion falls below the sensitivity limits for the target, the companion is not detected. \\
At the end of the N runs of the artificial survey, we know how many planets or BDs we have ``created'' and how many we would have detected in each simulation. Finally, we can define the total probability for the survey of having found zero, one or more objects, given the model that we assumed, and can compare it with the real outcome of the survey.
Examples of applications of this tool are presented in Section~\ref{sect:3.5}. Here we show the results obtained by using the COND evolutionary models. \cite{Dupuy2014} present evidence for a substellar luminosity problem in the predictions from evolutionary models. The predicted luminosities from which we estimate planet and BD masses could in fact be underestimated \citep{Dupuy2014}, which would make our survey results more constraining regardless the frequency of planetary mass objects.  Although cloudless, the COND models show the smallest discrepancy with the observed luminosities compared to the DUSTY models \citep{Chabrier2000} and the \cite{Saumon2008} hybrid models. The DUSTY models might be more appropriate for some of the higher mass (and presumably hotter) brown dwarfs that we simulate. However the simulations with both families of evolutionary models lead to consistent results.

\section{Dataset}\label{sect:3.4}
The dataset that we used to test the model for the substellar CMF described in Section~\ref{sect:3.2} consisted of the targets of the NaCo-LP \citep{Chauvin2015,Desidera2015}, a large observing program (ESO: 184.C-0157) in the context of the VLT/SPHERE scientific preparation.
This program is a direct imaging survey providing a homogeneous statistically significant study of the occurrence of giant planets and BDs in wide (5-500 AU) orbits around young, nearby stars.
It focused on a carefully selected sample, chosen with declinations $\delta \leq 25^{\circ}$, ages $t\leq$ 200 Myr, distances $d\leq$100 pc, $R$-band brightnesses $R \leq 9.5$ and with exclusion of spectroscopic binaries and visual binaries with separation $<$6''.
The sample was initially comprised of 84 targets, none of which had been observed before in a planet survey. These stars were then observed with VLT/NaCo in the $H$ band  between end 2009 and 2013, for a total of 16.5 nights. A complete description of the sample and its properties (distance, age, mass) can be found in \cite{Desidera2015}. The summary of the program and of the observations is provided in \cite{Chauvin2015}. As a result of the campaign, no substellar companions were detected around the targets. For 51 targets the observations were complete up to 300 AU and they constitute the statistical sample presented in section 6.1 of \cite{Chauvin2015}.  Once only FGK stars are considered,  the sample reduces to 47 targets. 
In the present work we used these 47 objects to place some constraints on the BD and planet populations together. For simplicity we will refer to the sample as ``NaCo-LP observed sample''.
The statistical analysis on this sample regarding the constraints on the planet distributions only is presented in a separate paper (Vigan et al., in preparation). 

 We complemented the observed sample with similar targets observed in previous surveys (not re-observed within the NaCo-LP) for which reduced data and detection limits were available:
\cite{Lowrance2005}, \cite{Masciadri2005}, \cite{Biller2007},  \cite{Kasper2007}, \cite{Lafreniere2007}, \cite{Chauvin2010}, \cite{Heinze2010}, \cite{Vigan2012}, \cite{Rameau2013}, and \cite{Brandt2014a}. 
We included in the so-called ``full sample'' targets that share the same range of spectral types (from early-F to late-K) and distance ($\leq 100$ pc) as the NaCo-LP stars, for a total of 152 objects.
No age or declination limit was applied in this case, while we adopted the same selection criteria as for NaCo-LP  targets concerning binarity. 
Ages and other stellar parameters were redetermined for the full sample using the procedures described in \cite{Desidera2015} for the NaCo-LP sample.
Full details on the revised stellar properties will be provided in Vigan et al., in preparation. \\
While the outcome of the NaCo-LP was a null result, substellar companions have been found around 4 targets in the archive sample::
\begin{description}
\item[GSC 08047-00232 B] is a 25$\pm$10 $M_{Jupiter}$ mass BD with a derived spectral type M9.5$\pm$1 \citep{Chauvin2004} at a projected separation of 278 AU. It is a probable member of the Tucana-Horologium association with an age of 10-50 Myrs.
\item[AB Pic b] is a $\sim13$ $M_{Jupiter}$ mass object at the planet/brown dwarf boundary \citep{Chauvin2005} and separation of 275 AU. It is also member of the Tucana-Horologium association.
\item[HD 130948 BC] is a BD binary system, companion to HD 130948 \citep{Potter2002}. The total mass was estimated to be 0.1095$\pm$0.0022 $M_{\odot}$ \citep{Dupuy2011} and the age 0.93 Gyr.
\item[PZ Telescopii B] is a BD companion with a mass of 24-40 $M_{Jupiter}$ at a projected separation of $\sim$15 AU \citep{Biller2010,Mugrauer2010}. It is a member of the $\beta$ Pic moving group.
\end{description}

The full sample (NaCo-LP $+$ archive) is therefore comprised of 199 nearby ($d\leq$100 pc) solar-mass stars that have been observed in deep imaging with sensitivity down to planetary mass companions.

\section{Monte Carlo simulation results}\label{sect:3.5}
To test our model for the substellar CMF, we used the tool described in Section~\ref{sect:3.3} to compare our predictions with the set of data presented in Section~\ref{sect:3.4}, beginning with the  NaCo-LP observed sample and then with the full sample. 

   \begin{figure*}[H]
   \resizebox{\hsize}{!}{
            \includegraphics[width=\textwidth]{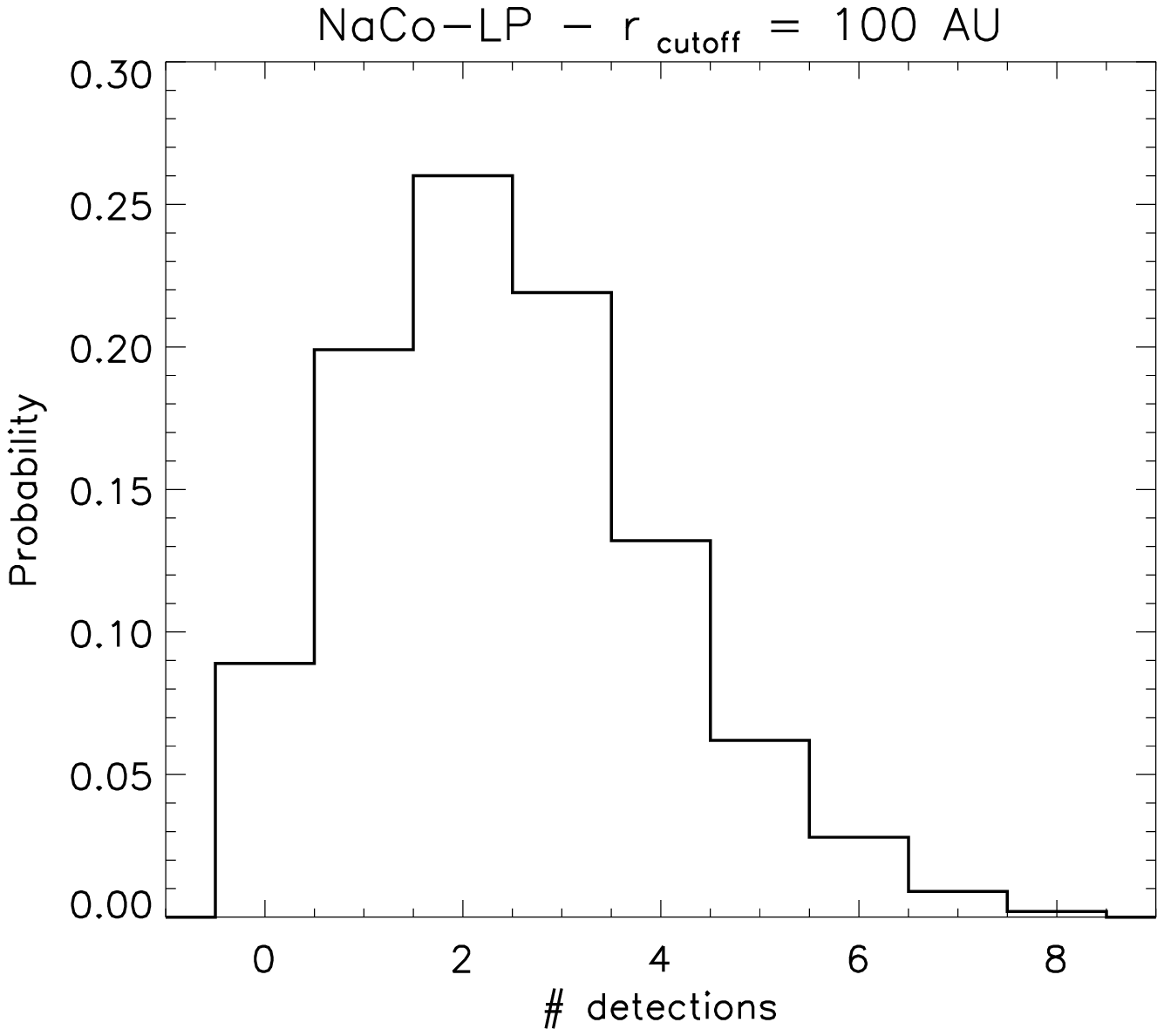}
            \includegraphics[width=\textwidth]{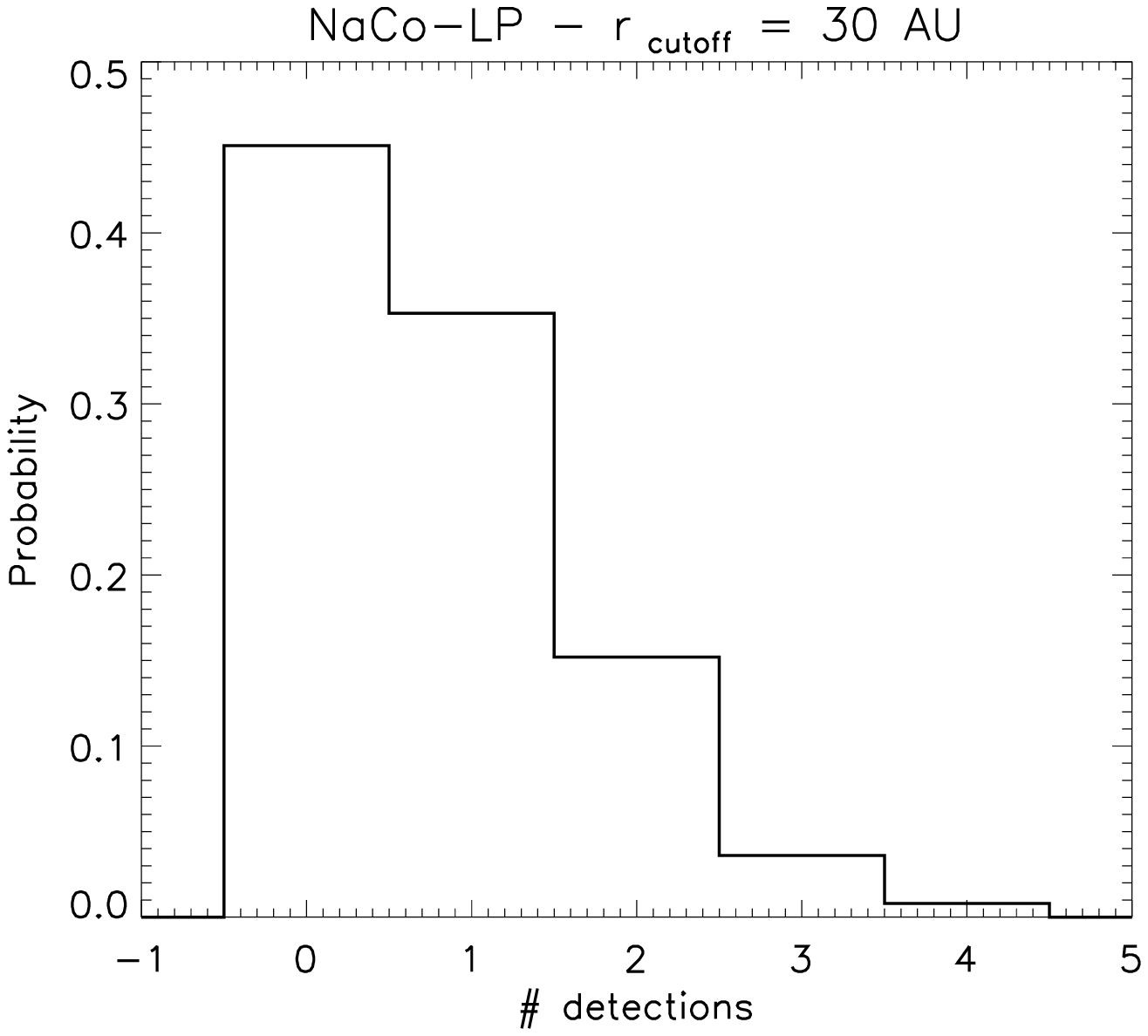}
            \includegraphics[width=\textwidth]{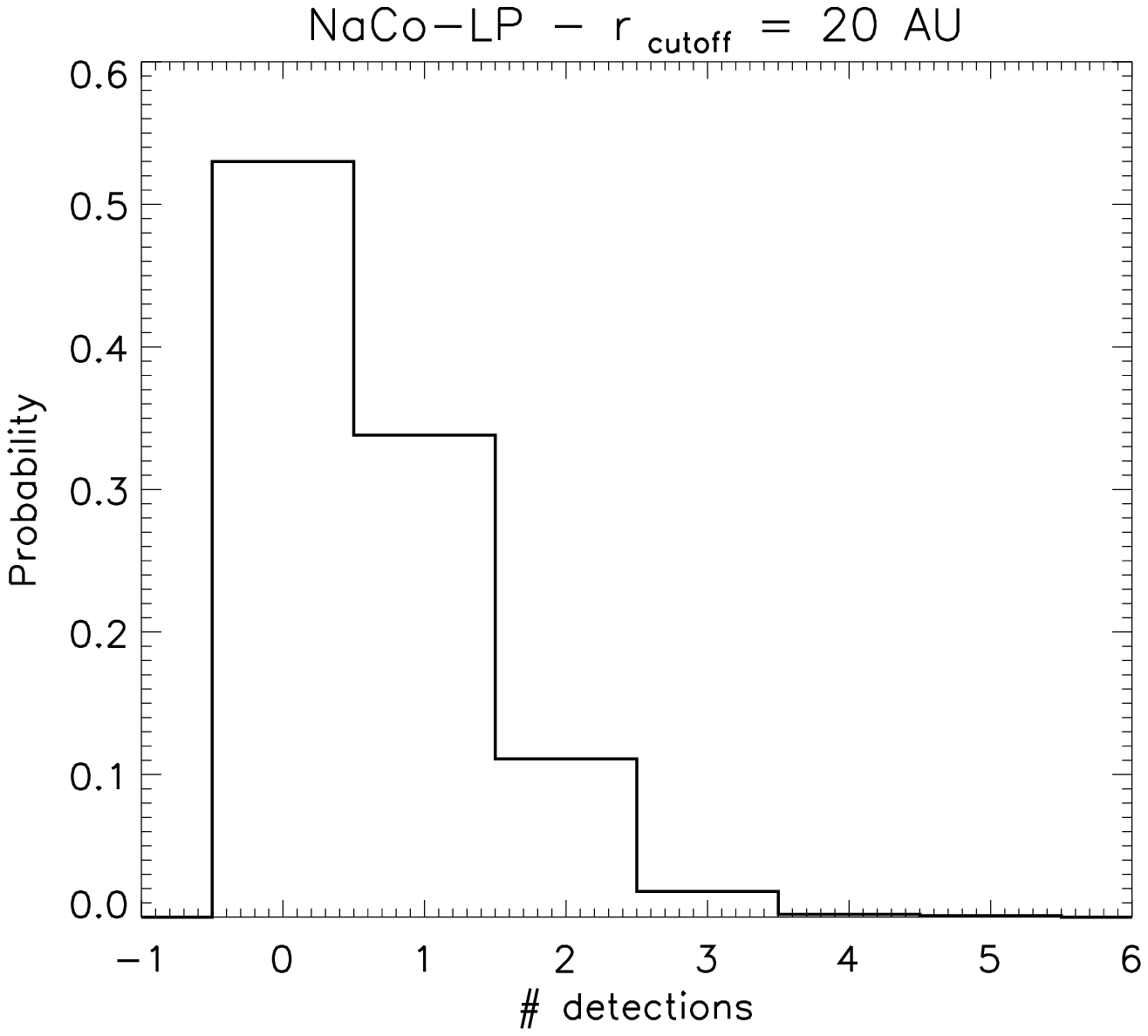}}
      \caption[Detection probability for the NaCo-LP observed sample]{Detection probability for the NaCo-LP observed sample for r$_{cutoff}$=100, 30, 20 AU (from left to right).
              }
         \label{figure:3.2}
   \end{figure*}

\subsection{Results from the NaCo-LP observed sample}\label{sect:3.5.1}
Concerning the 47 targets observed within the NaCo-LP,  we initially ran 3000 Monte Carlo simulations of the survey, with the set of parameters shown in Table~\ref{table:3.1} and with an outer radius cutoff for the planet separation distribution of r$_{cutoff}$=100 AU to quantify how likely the null result of the survey would be, according to this model.
In the 3000 simulations 410847 planets and 3714 BDs were created, of which 5856 and 1527 were detected, respectively. The overall probability distribution of detections is shown in Figure~\ref{figure:3.2}(a). In 9\% of the realizations, our survey found zero companions, while in 19\% of the cases it found one and 71\% of the time it found two or more planets or BDs. 
We then verified the dependence of this result on the choice of r$_{cutoff}$. We repeated the set of 3000 Monte Carlo simulations for both r$_{cutoff}$=20 AU and r$_{cutoff}$=30 AU, but leaving all the other parameters unchanged. In this case, the probability of a null results is 53\% and 45\% for r$_{cutoff}$=20 AU and 30 AU, respectively (see Figure~\ref{figure:3.2}). 
This result indicates that introducing a lower value for the outer radius cutoff better reproduces the outcome of the observations, assuming that the mass and SMA distributions from RV surveys \citep{Cumming2008} are a good representation even at larger separations.
However, the complete sample is needed in order to put more stringent constraints.

\subsection{Results from the full sample}\label{sect:3.5.2}
In the case of the full sample (199 targets), we repeated the same sets of simulations that we carried out for the NaCo-LP observed dataset.
Besides the size, the main difference between the NaCo-LP subsample and the complete dataset is in the outcome of the observations. 
   \begin{figure*}[H]
      \resizebox{\hsize}{!}{
            \includegraphics[width=\textwidth]{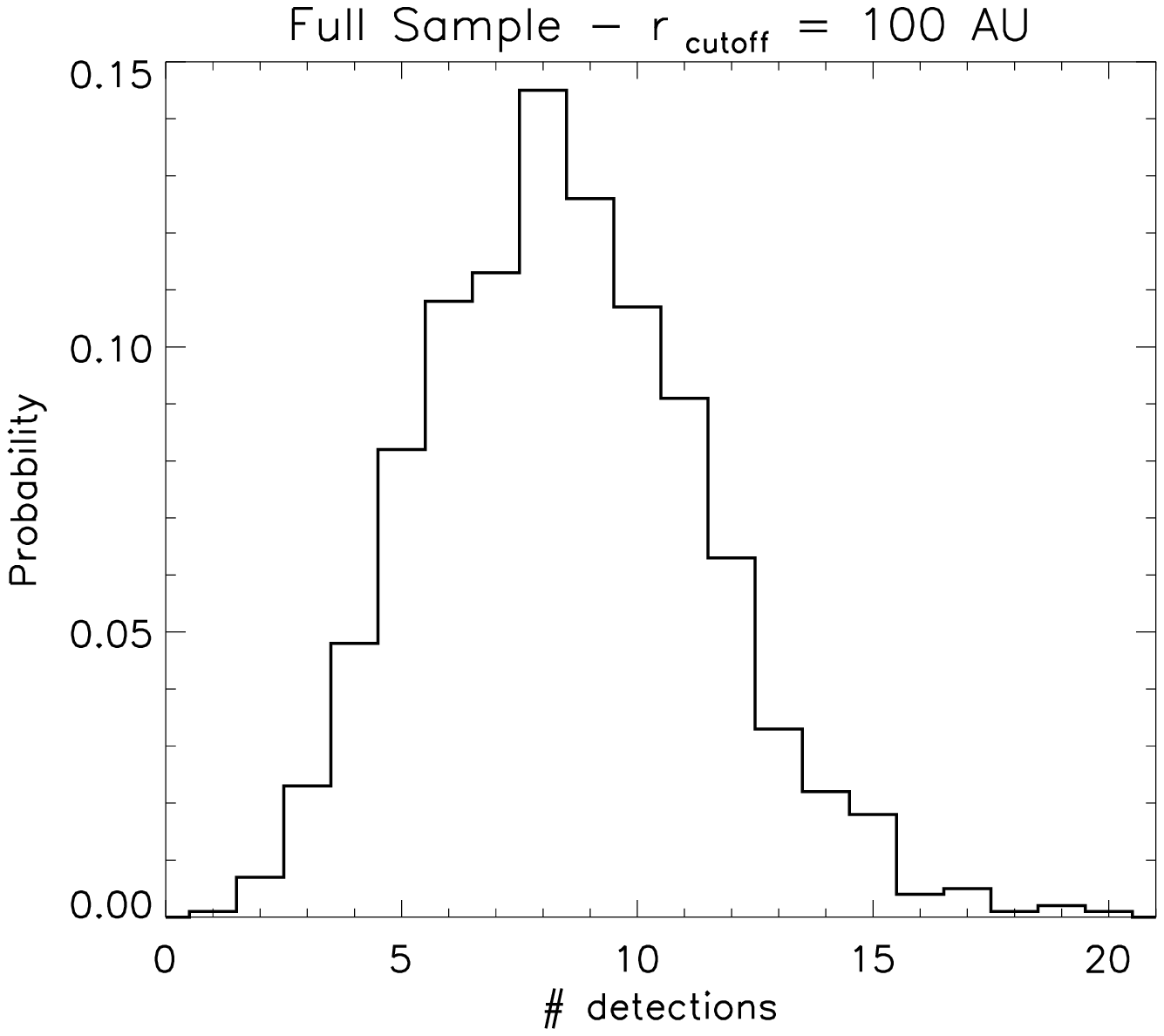}
            \includegraphics[width=\textwidth]{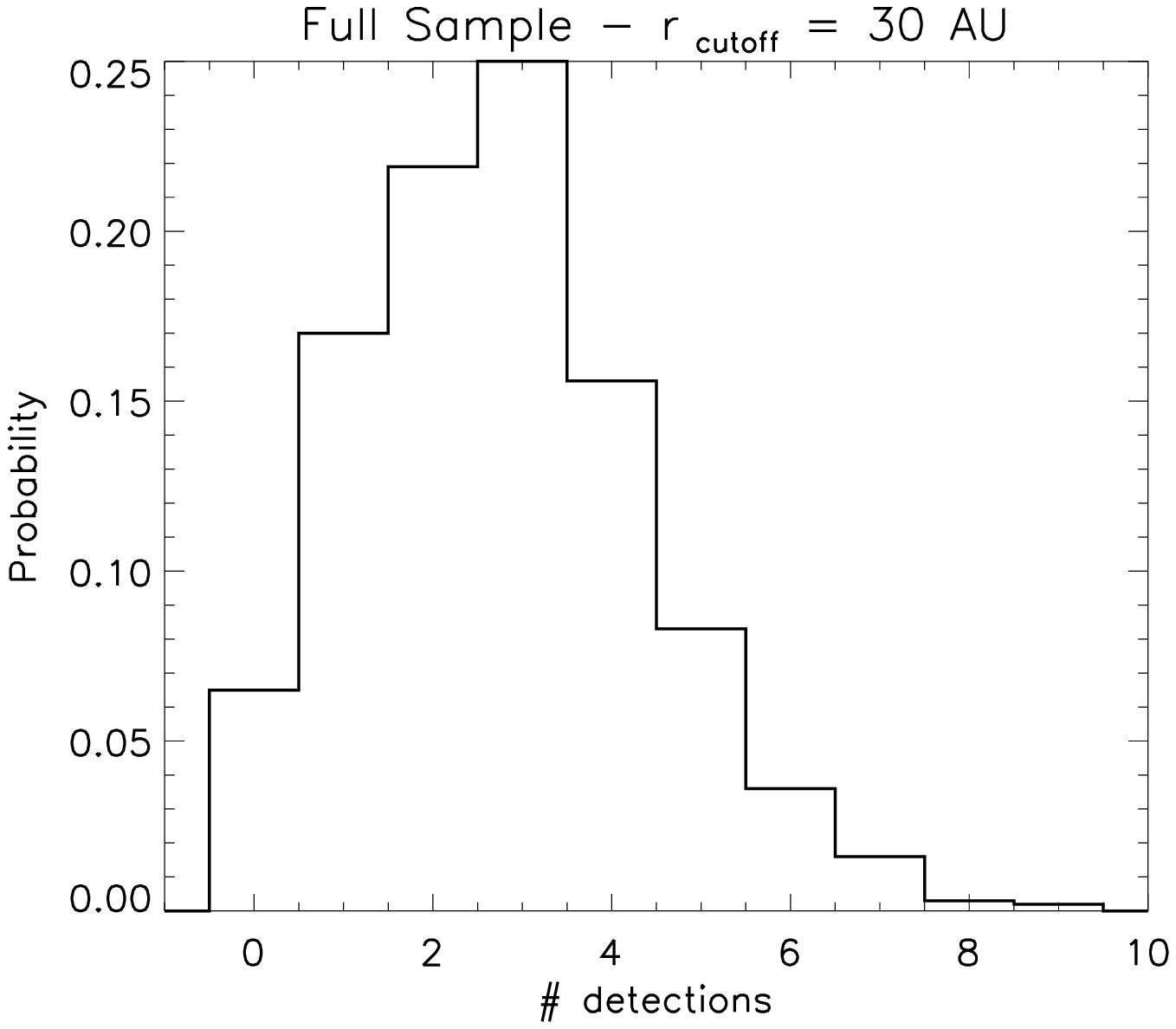}
            \includegraphics[width=\textwidth]{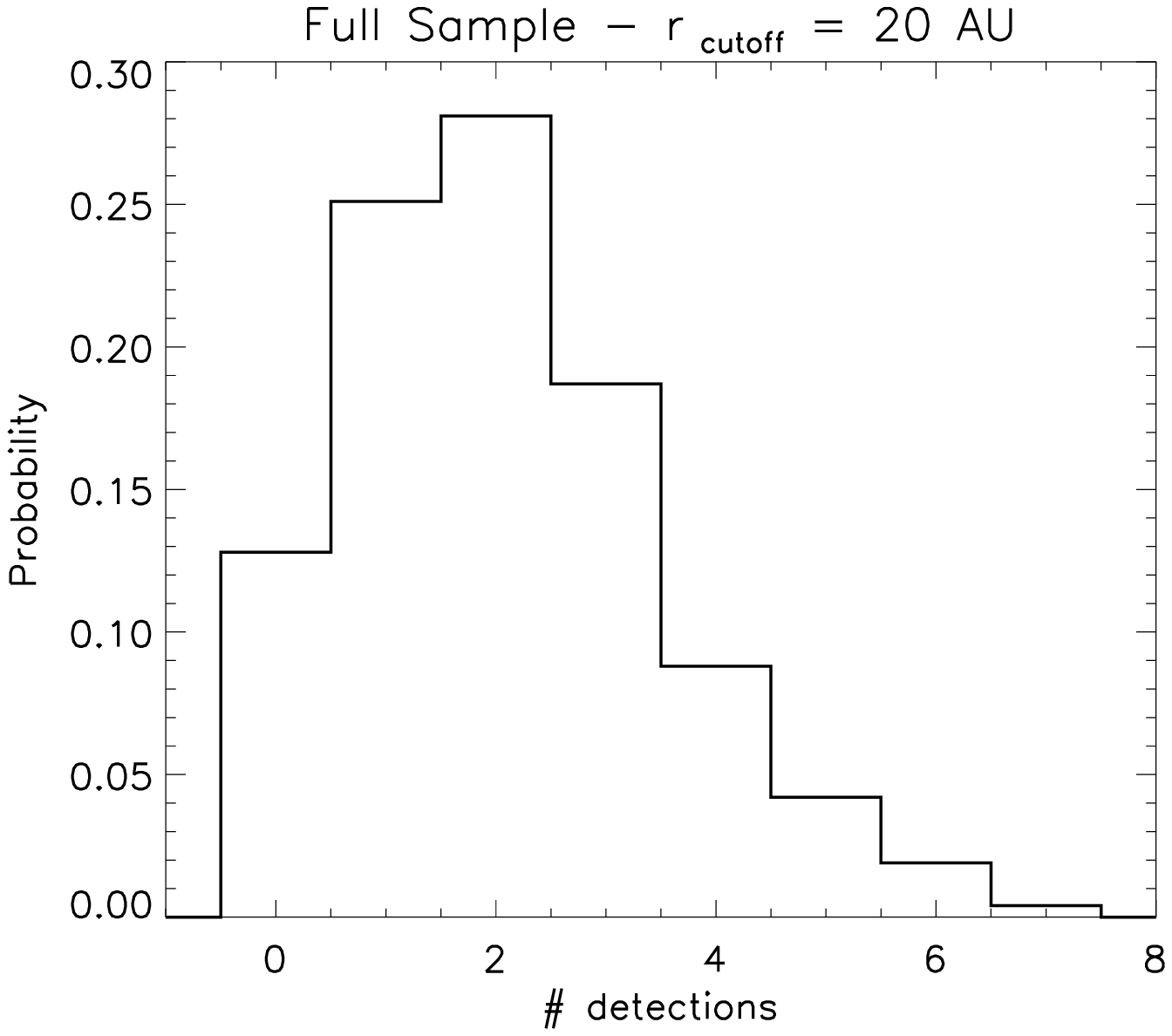}}
      \caption[Detection probability for the full sample]{Detection probability for the full sample for r$_{cutoff}=$100, 30, 20 AU (from left to right). }
         \label{figure:3.3}
   \end{figure*}
In this case, because of the detection of the BD companions listed in Section~\ref{sect:3.4}, we are interested in the probability of detecting 3 substellar objects. We note that we do not consider the tight BD binary HD 130948 BC among the detections. It constitutes in fact a hierarchical triple system with the sun-like star HD 130948 A. Such a system would generally contribute to the CMRD as two different binaries with masses $M_A$ and $M_{BC}$, where $M_{BC}=M_{B}+M_C$, and  $M_{B}$ and $M_C$, respectively. For HD 130948, in both binary systems either the primary or the secondary components would be outside the mass ranges of interest of this study. 

Analogously to what has been done for the NaCo-LP observed sample, we simulated 3000 artificial observations of the targets in the full sample for the same three values of r$_{cutoff}$= 20,30 and 100 AU. 
The detection probability distributions are shown in Figure~\ref{figure:3.3}.
The probability of detecting 3 objects is 2\%, 25\% and 19\% for r$_{cutoff}$= 100, 30 and 20 AU, respectively. The results from the full sample are similar to what was obtained for the NaCo-LP dataset only.
The model described in Section~\ref{sect:3.2} cannot be ruled out by the observations and smaller truncation radii give a better agreement.

\begin{figure}
               \resizebox{\hsize}{!}{\includegraphics{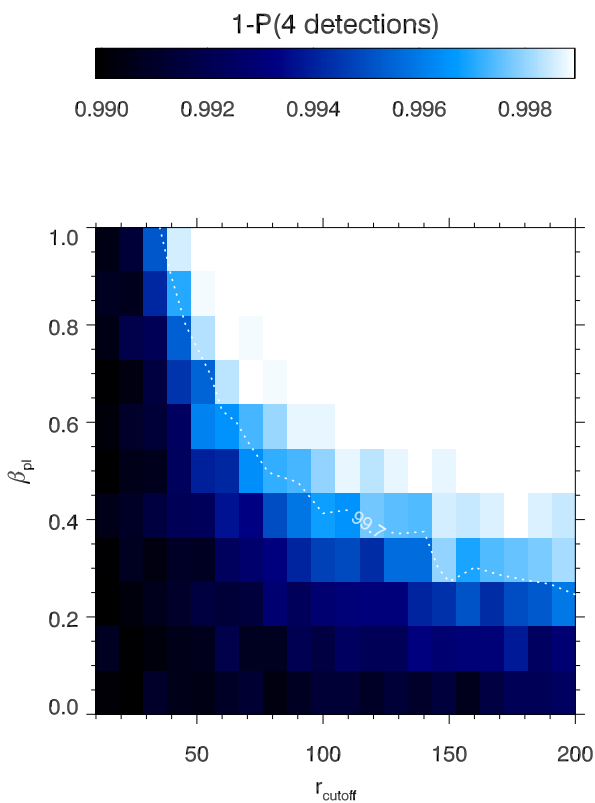}}
      \caption[$\beta_{pl}$-r$_{cutoff}$ parameter space]{Probability of our surveys detecting 3 substellar companions, as a function of the power-law slope of the semi-major axis distribution $\beta_{pl}$, and the outer radius cutoff of the semimajor axis distribution.  The overall probability is normalized to be 1 within the figure. The region above the line can be ruled out at 99.7\% confidence.}
         \label{figure:3.4}
   \end{figure}

In order to quantify this statement and assess which planet distributions could instead be excluded, we explored the SMA power-law index ($\beta_{pl}$) vs. outer truncation radius (r$_{cutoff}$) parameter space. Having fixed the BD mass and semi-major axis distributions, we varied $\beta_{pl}$ from 0 to 1, with steps of 0.1, and r$_{cutoff}$ from 10 to 200 AU with steps of 10 AU. For each pair of $\beta_{pl}$ and r$_{cutoff}$, we ran 300 simulations of the survey and we calculated the probability of detecting up to 3 objects. Regardless of the choice of values for $\beta_{pl}$ and r$_{cutoff}$, every simulation was normalized to match the RV statistics within 3 AU.
Figure~\ref{figure:3.4} shows the probability of detecting up to 3 substellar companions as a
function of  $\beta_{pl}$ and r$_{cutoff}$. Each grid cell point represents the 300 Monte Carlo simulations of the survey. In order to estimate confidence levels, the overall probability is normalized to be 1 within the figure. The confidence level shown in Figure~\ref{figure:3.4} indicates where the probability is lower than  0.3\%. The region above the line therefore can be ruled out at a 99.7\% confidence. For example, in the case of $\beta_{pl}$=0.39, as suggested by RV measurement \citep{Cumming2008}, all models with r$_{cutoff}\gtrsim$100 AU can be ruled out at a $>3 \sigma$ confidence.

\section{Discussion}\label{sect:3.6}
The results presented in Section~\ref{sect:3.5} show that the current measurements of the occurrence of planets and BDs in wide orbits around solar-type stars do not rule out the model for the substellar CMF, presented in Section~\ref{sect:3.2}. The Kolmogorov-Smirnov test (KS test) between the masses of the detected BDs and the proposed CMF returns a probability $P_{KS}>45$\%, for all choices of truncation radii. Such probability does not reject the null hypothesis that they come from the same parent population. If the RV statistics \citep{Cumming2008} provide a good estimate of the normalization and the power-law slopes of the planet distributions, an outer truncation radius of at most 100 AU is necessary to reproduce the observations (at a 99.7\% confidence). These constraints on the planet population are consistent with previous survey analyses \citep[][]{Biller2007,Nielsen2010,Chauvin2010}.

In the past years, the observed dearth of close ($\leq$ 5 AU) BD companions to solar-mass stars \citep[e.g.][]{Marcy2000}, also known as the ``brown dwarf desert'', has been used as evidence for a separate formation mechanism for brown dwarfs with respect to stars. \cite{Grether2006}, in an attempt to quantify the relative number of stellar companions, planets and BDs in close orbit (period $\leq$ 5 yrs) around nearby sun-like stars, found a paucity of objects in the BD mass regime compared to planetary and stellar companions, with the driest part of the ``desert'' being at 31$^{+25}_{-18}$ $M_{Jupiter}$. \cite{Sahlmann2011} also measured the CMF for close BD companions to solar-type stars and found a minimum in the distribution between 25 and 45 $M_{Jupiter}$ and claimed to have detected the high mass tail of the planet mass distribution.

Based on our analysis, the mass distribution of substellar companions is consistent with a superposition of the planet CMF and the stellar CMRD extrapolated into the BD mass regime provided that the planet distribution is truncated at a few tens of AU. This suggests that objects from the Hydrogen burning limit down to a few Jupiter masses may still form as stellar companions, without the need of introducing a separate formation mechanism. From the theoretical point of view, recent turbulent fragmentation models have also explained the binary brown dwarf properties in terms of a single core fragmentation mechanism \citep{Jumper2013}. The paucity of companions in the mass range 10-50 $M_{Jupiter}$ even at large ($>$10 AU) separations would naturally arise from the superposition of the two mass distributions (see Figure~\ref{figure:3.1}). 
On the observational side, it is worth mentioning that some planetary-mass companions identified by direct imaging in young clusters show evidence for active accretion and are surrounded by massive disks \citep[see e.g.,][]{Seifahrt2007,Bowler2011,Joergens2013,Zhou2014}. This also argues for an extension of the stellar/BD formation mechanisms in the planetary-mass regime.
Our results are also consistent with the conclusions of \cite{Brandt2014b}, that some of the so-far directly imaged exoplanets could have formed by either core or disk fragmentation and constitute the low-mass tail of the brown dwarf mass distribution.

In the next few years, new instruments, like the VLT Spectro-Polarimetric High-contrast Exoplanet REsearch (VLT/SPHERE) or the Gemini Planet Imager (GPI), will enable us to find new planetary-mass companions. To what extent large surveys with these new instruments will help us constrain the distributions of orbital parameters still remains to be seen. With the next generation of extremely large telescopes instead (e.g. the European Extremely Large Telescope, E-ELT), our understanding of planet statistics will make a huge step forward. It will be possible to directly measure the shape of the substellar CMF and locate with higher precision the minimum even for wide companions. In this context, it will be interesting to study how the distribution of companion masses in the substellar regime varies not only with respect to separation but also as a function of primary mass.

\section{Conclusions}\label{sect:3.7}
In this paper we propose a simple model for the substellar mass spectrum, as a combination of the planet CMF and an extrapolation of the stellar CMRD into the BD mass regime. 
Taking advantage of the largest sample to date of solar-type stars observed with direct imaging, we ran Monte Carlo simulations to compare predictions of our model with the observations from the NaCo-LP and archival data.
We conclude that:
\begin{itemize}
\item The outcome of the direct imaging surveys is consistent with a superposition of the CMF derived by RV measurements ($\alpha_{pl}=-0.31$ and $\beta_{pl}=0.39$) and of the stellar CMRD down to 5 $M_{Jupiter}$, as long as r$_{cutoff} \lesssim$ 100 AU  (99.7\% confidence).
\item When all the other parameters are fixed, some combinations of $\beta_{pl}$-r$_{cutoff}$ can be ruled out by the observations with a 99.7\% confidence.
\item The proposed CMF has a minimum between 10-50 $M_{Jupiter}$, in agreement with the results from RV observations \citep{Sahlmann2011}. The KS test probability between this CMF and the masses of the 3 detections is $P_{KS}>45$\%, and does not reject the hypothesis that the data were drawn from the same distribution as the model.
\item In this picture the so-called ``BD desert'' would naturally arise from the shape of the mass distribution, without having to introduce any different formation mechanism for BDs.
\end{itemize}

Future observations may allow us to measure directly the shape of the distribution and the precise location of the minimum in the substellar mass spectrum. 

\begin{acknowledgements}
We are grateful to the referee for the useful comments. Part of this work has been carried out within the frame of the National Centre for Competence in Research PlanetS supported by the Swiss National Science Foundation (SNSF). MR, SPQ, EB, JH and MRM acknowledge the financial support of the SNSF. JC is supported by the U.S. National Science Foundation under Award No. 1009203. SD, ALM, RG, and DM acknowledge partial support from PRIN-INAF 2010 ``Planetary systems at young ages and the interactions with their active host stars''. A. Z. acknowledges support from the Millennium Science Initiative (Chilean Ministry of Economy), through grant ``Nucleus RC130007''.
\end{acknowledgements}


\end{document}